\documentclass[  conference]{IEEEtran}
\usepackage{cite}
\usepackage{amsmath,amssymb,amsfonts}
\usepackage{graphicx}
\usepackage{textcomp}
\usepackage{xcolor}
\usepackage[hyphens]{url}
\usepackage[caption=false]{subfig}
\usepackage{multirow}
\usepackage{arydshln}
\usepackage[linesnumbered,ruled,vlined]{algorithm2e}
\usepackage{pgfplots}
\usepackage{svg}

\def\BibTeX{{\rm B\kern-.05em{\sc i\kern-.025em b}\kern-.08em
    T\kern-.1667em\lower.7ex\hbox{E}\kern-.125emX}}

\newcommand*\circled[1]{\tikz[baseline=(char.base)]{
    \node[shape=circle,draw,inner sep=1pt] (char) {#1};}}

\usepackage[noend]{algpseudocode}
\algnewcommand{\LeftComment}[1]{\(\triangleright\)#1} 

\title{Forecasting lifetime and performance of a novel NVM last-level cache with compression}

\author{\IEEEauthorblockN{Carlos Escuin\IEEEauthorrefmark{1}, Pablo Ibañez\IEEEauthorrefmark{1}, Teresa Monreal\IEEEauthorrefmark{2}, Jose M. Llaberia\IEEEauthorrefmark{2}, Victor Viñals\IEEEauthorrefmark{1}}
\IEEEauthorblockA{\IEEEauthorrefmark{1}University of Zaragoza}
\IEEEauthorblockA{\IEEEauthorrefmark{2}Universitat Politècnica de Catalunya}}

\begin{document}
\maketitle
\thispagestyle{plain}
\pagestyle{plain}


\section{Introduction}


The goal of the last-level cache (LLC) in the memory subsystem of a chip multiprocessor is to filter main memory requests from the private levels and service them with a lower latency.
The number of integrated cores on a chip is increasing and so it is the LLC size.
Most of LLCs are built using SRAM technology that does not scale well in terms of density and static power.
Given this limitation, non-volatile (NV) memories are interesting alternatives to replace on-chip SRAM due to their higher density and lower static power~\cite{Korgaonkar-18-DensityTradeoffs-WCAB-VHC}.
However, write operations are costly in terms of latency and energy, and gradually wear out the bitcells until a hard fault occurs.
Compared to SRAM, the write endurance in NV memories is much lower and is usually modeled by a normal distribution of mean $10^w$ write operations, $w$ depending on technology, manufacturer and target market~\cite{Schechter-10-UseECPnotECC, Natsui-20-DualPortSOT-MRAM, Farbeh-16-FloatingECC}.

Several works address the write endurance problem in complementary ways such as evenly distributing writes across the whole memory structure (wear-leveling~\cite{Wang-13-I2WAP, Farbeh-16-FloatingECC}), reducing the amount of written information~\cite{Dgien-14-CompressionArchitecture}, or using redundant storage~\cite{Schechter-10-UseECPnotECC}.
In order to cope with hard faults, memory structures must include a way to detect and correct them, generally with error correction codes, ECCs.
A bitcell failure requires deactivating the corresponding memory region it belongs to: a whole memory page~\cite{Schechter-10-UseECPnotECC}, cache frame~\cite{Chang-07-BD1}, or single byte~\cite{Ferreron-15-Concertina}.

In summary, to improve the state of the art in non-volatile LLCs (NV-LLC), proposals are needed that simultaneously reduce the number of writes, maintain uniform wear on all bit cells, and continue to operate even if some of the rated capacity is out of service.
Thus, our \textbf{first contribution} is a novel NV-LLC design which combines byte-level disabling and compression.
It can include any compression mechanism offering low decompression latency, high coverage and sufficient compression ratio.
We select Base-Delta-Immediate (BDI) compression, which is well positioned in this tradeoff~\cite{Pekhimenko-12-BaseDeltaImmediate}.

Cache block compression is used to achieve two effects, namely to reduce the amount of information written (to increase the cache lifetime) and to allow compressed blocks to be written to degraded cache frames (to increase the cache hit rate).
To this end, we propose a circuitry that rearranges the bytes of a compressed block among the healthy bytes of a cache frame, while also guaranteeing uniform wear. A full-custom VLSI design demonstrates its feasibility in terms of area, latency and power consumption.

Previous work assesses lifetime improvement indirectly through relative reduction in the number of writes~\cite{Wang-13-I2WAP}, or by using approximate aging models in the context of implementing main memory with NV technologies~\cite{Schechter-10-UseECPnotECC}.
To the best of our knowledge, there is no model able to predict the time evolution of the capacity or performance of a NV-LLC that degrades as a consequence of writes.
Therefore, our \textbf{second contribution} is such a forecasting procedure, suitable for multicore NV-LLC with or without compression. It proceeds through successive epochs. Within each epoch, a cycle-accurate simulation and byte failure prediction is performed.

\section{Compression-based NV-LLC}
\label{sec:nvllc}


We assume a SECDED mechanism, capable of correcting a single-bit fault and detecting up to two. The mechanism, upon detecting and correcting a single bit fault, triggers an operating system exception. In the absence of compression, the exception service routine has no choice but to disable the entire cache frame to avoid a second fault which could no longer be corrected.
The baseline systems considered below are state-of-the-art proposals that implement this coarse-grained disabling~\cite{Chang-07-BD1, Schechter-10-UseECPnotECC}. On the contrary, our proposal allows disabling at much finer byte granularity, taking advantage of compression to store cache blocks of reduced size.

To slow aging, it is necessary to avoid  write concentration in specific regions of the cache. The conventional solution seeks an even distribution of writes among cache sets, and for this we index sets using a good hash function.
However, since compression allows only part of the bytes of a frame to be written, our proposal also needs to ensure a balanced wear across all the bytes of the frame, and even when the capacity of the frames decreases. Therefore, we propose a new intra-line wear leveling mechanism based on byte rearrangement and a global counter that points out the starting point of frame writing.

\begin{figure}[ht]
    \centering
    \includegraphics[width=0.95\linewidth]{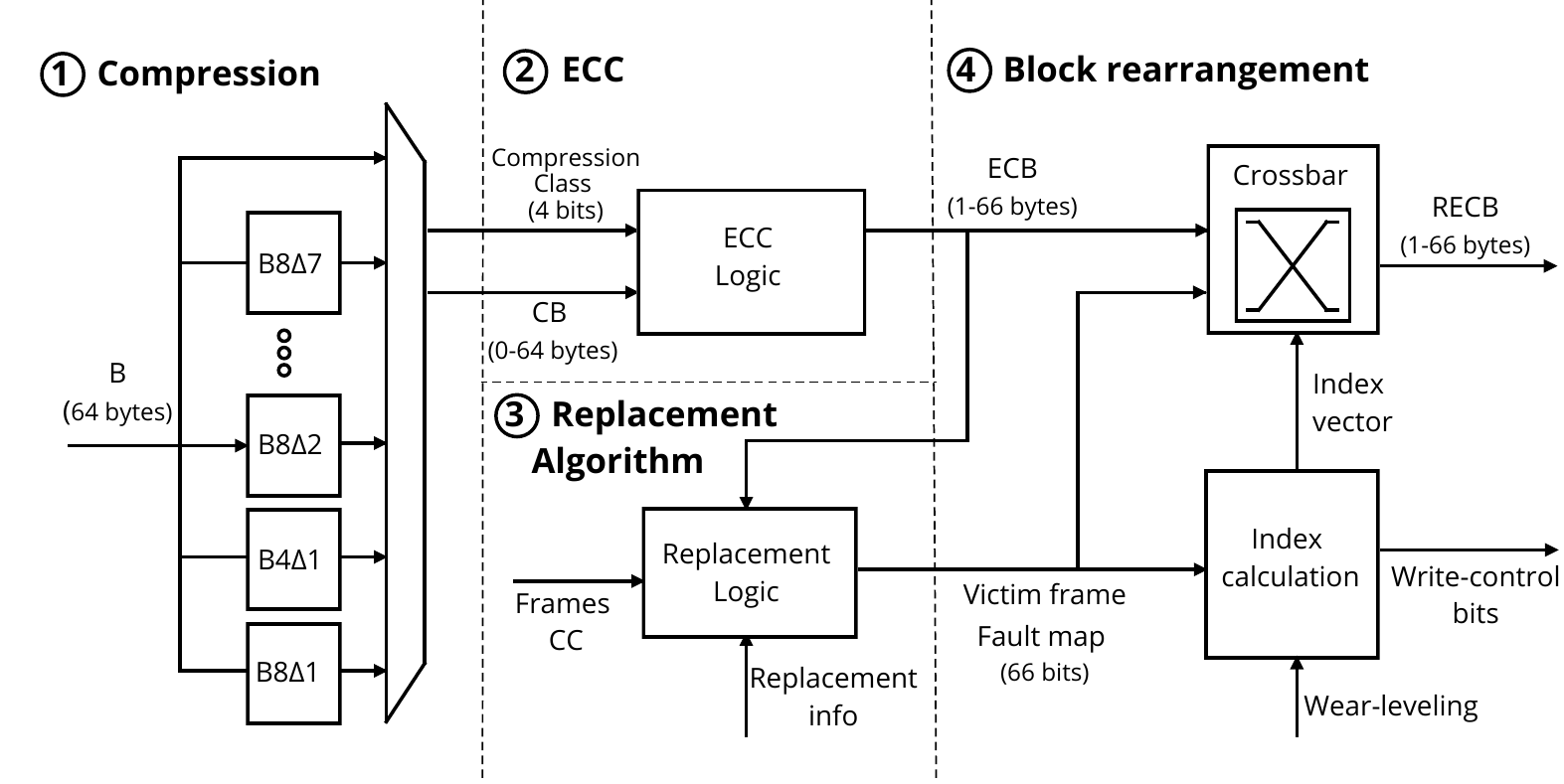}
    \caption{Writing compressed blocks to the NV-LLC.}
    \label{fig:nvllc}
\end{figure}

Figure \ref{fig:nvllc} explains the steps taken in a block write. First, the compression unit receives the block B, \circled{1} Compression.
The result of each BDI compressor is: a) whether the compression is feasible and, if so, b) the compressed block, CB.
Next, the corresponding ECC bits are computed from CB, \circled{2} ECC.
The length of the compressed block and the ECC bits (ECB) determine the minimum capacity a frame must have to accommodate it.
Accordingly, the replacement algorithm selects the victim block from the subset of frames with the required minimum capacity, \circled{3} Replacement Algorithm.
Every frame has an associated bitmap that points out the faulty bytes.
Finally, in relation to the intra-frame wear-leveling mechanism, a global counter points out the initial write position of the block within the frame, and from it the block is rearranged for selective writing (RECB) and the write control bits  are generated, \circled{4} Block rearrangement.

\vspace{-0.15cm}
\section{Forecasting procedure}
\label{sec:forecast}

To forecast the evolution of the effective capacity over time, we propose to initialize a map with the number of remaining writes of each cache byte (RW map) using a statistical model of resilience of the selected memory technology~\cite{Farbeh-16-FloatingECC, Schechter-10-UseECPnotECC}.
A naive, but exact, approach is to simulate the progressive degradation of the NV-LLC: each time a write occurs, it is reflected in the RW map by degrading the corresponding bytes. When a byte suffers a hard fault it is deactivated and the simulation continues with the system a little more degraded. However, to get realistic results, the simulation must be cycle-accurate and driven by a realistic workload, but in doing so we could only forecast time lapses of a few milliseconds.

\begin{figure}[ht]
    \centering
    \includegraphics[width=\linewidth]{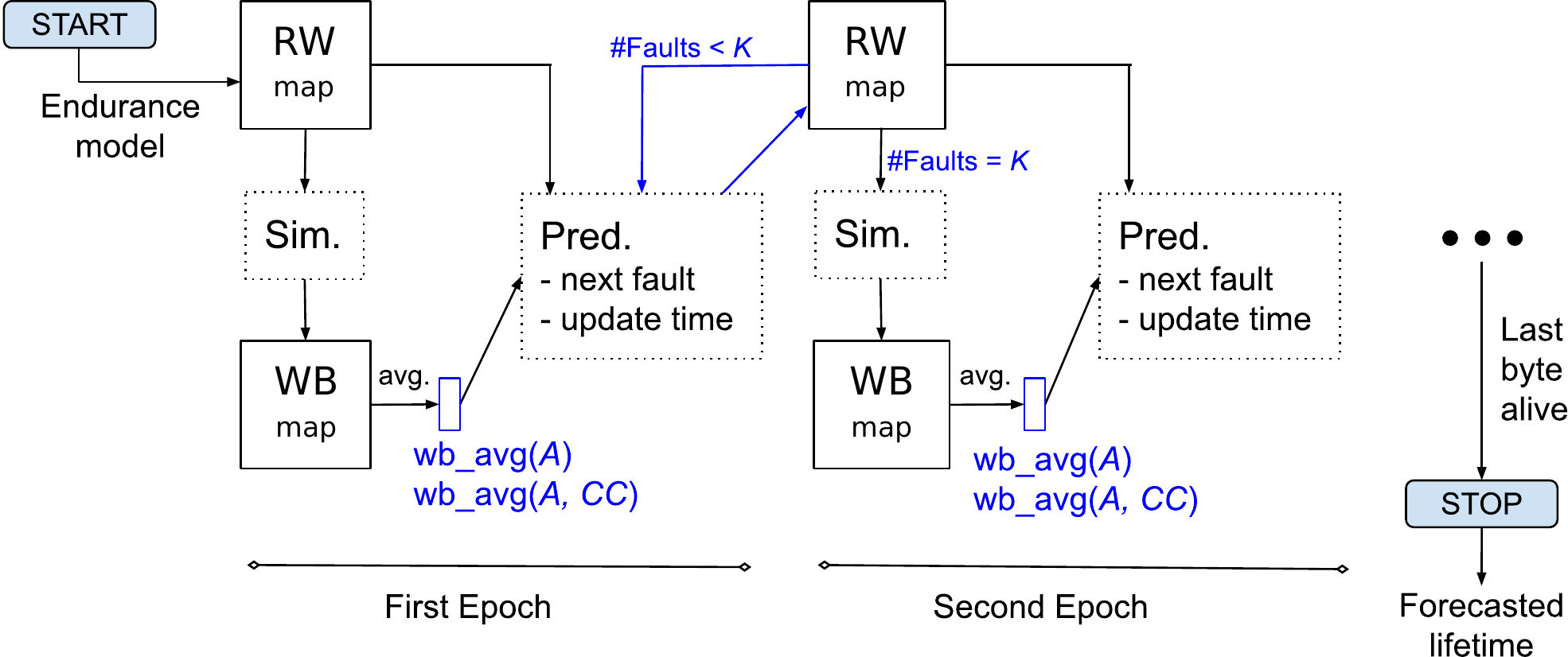}
    \caption{Forecasting procedure flow diagram.}
    \label{fig:forecast}
\end{figure}

An approximate approach is to move faster in time by simulation-prediction epochs, see Figure~\ref{fig:forecast}. At each epoch, a short simulation calculates the \textit{number of writes per unit time to each byte}, recording them in a \textit{per-byte write bandwidth} map (WB map). At the end of the simulation it is calculated which byte will die next and in which time, updating the RW map with that byte deactivated.
This approach needs as many simulations as the number of bytes in the LLC, which again means an unaffordable simulation time.

To decrease the number of simulations, we propose to approximate a little further by making \textit{K} consecutive predictions from a single simulation, see  blue text in Figure~\ref{fig:forecast}.  For an uncompressed cache we compute the average WB in sets with \textit{A} alive frames, \textit{wb\_avg(A)}, applying this aging rate to the RW map: to each byte, \textit{wb\_avg(A)} is applied if its frame is in a set with \textit{A} alive frames. The procedure can be extended to the cache with compression also taking into account the compression class (CC) each frame belongs to and calculating new averages \textit{wb\_avg(A,CC)}.

\section{Evaluation}
\label{sec:evaluation}

A multicore processor with the latency parameters of an  STT-RAM LLC was simulated using the gem5 full-system simulator. The workload consists of application mixes taken from SPEC CPU 2006 and 2017. The bitcells write endurance follow a normal distribution of $\mu=10^{11}$ maximum number of write operations and $\sigma=0.2\mu$~\cite{Schechter-10-UseECPnotECC, Farbeh-16-FloatingECC}.

\begin{figure}[ht]
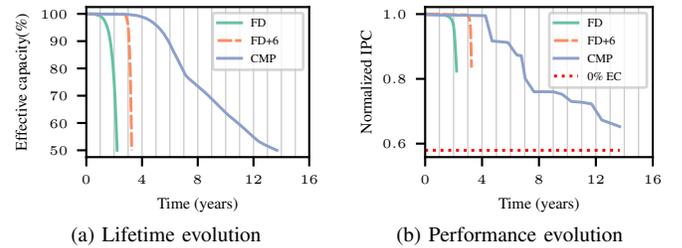

    \centering
    \subfloat[Lifetime evolution]{\label{subfig:lifetime_020}
        \input{Figuras/lifetime_nvmw.pgf}
    }
    \hfill
    \subfloat[Performance evolution]{\label{subfig:ipc_020}
        \input{Figuras/ipc_nvmw.pgf}
    }
    \caption{Lifetime and performance evolution until the NV-LLC reaches 50\% of the effective capacity in a 4-core on-chip multiprocessor.}
    \label{fig:evolution}
\end{figure}

Figure \ref{fig:evolution} shows the evolution of the system performance and capacity of a 4MB LLC with the above characteristics, from the beginning of operation until reaching 50\% of its effective capacity.
The byte-disabling compression-based system (CMP) is compared against two baseline systems:  Frame-Disabling (FD)~\cite{Chang-07-BD1} and ECP6 wich provides every frame with 6 ECPs (FD+6)~\cite{Schechter-10-UseECPnotECC}.
FD and FD+6 do not implement compression and disable  whole cache frames when a byte experience a fault.
Compared to FD, FD+6 and CMP multiply the time required to lose half the capacity by factors 1.47$\times$ and 6.2$\times$, respectively. CMP loses performance very gradually. For example, the time required to lose 15\% of IPC is 3.1 and 2.1 times longer in CMP than in FD and FD+6, respectively.

\vspace{-0.15cm}


\bibliographystyle{IEEEtranS}
\bibliography{refs}

\begin{thebibliography}{1}
\providecommand{\url}[1]{#1}
\csname url@samestyle\endcsname
\providecommand{\newblock}{\relax}
\providecommand{\bibinfo}[2]{#2}
\providecommand{\BIBentrySTDinterwordspacing}{\spaceskip=0pt\relax}
\providecommand{\BIBentryALTinterwordstretchfactor}{4}
\providecommand{\BIBentryALTinterwordspacing}{\spaceskip=\fontdimen2\font plus
\BIBentryALTinterwordstretchfactor\fontdimen3\font minus
  \fontdimen4\font\relax}
\providecommand{\BIBforeignlanguage}[2]{{%
\expandafter\ifx\csname l@#1\endcsname\relax
\typeout{** WARNING: IEEEtranS.bst: No hyphenation pattern has been}%
\typeout{** loaded for the language `#1'. Using the pattern for}%
\typeout{** the default language instead.}%
\else
\language=\csname l@#1\endcsname
\fi
#2}}
\providecommand{\BIBdecl}{\relax}
\BIBdecl

\bibitem{Chang-07-BD1}
J.~Chang \emph{et~al.}, ``The 65-nm 16-mb shared on-die l3 cache for the
  dual-core intel xeon processor 7100 series,'' \emph{JSSC}, 2007.

\bibitem{Dgien-14-CompressionArchitecture}
D.~B. Dgien \emph{et~al.}, ``Compression architecture for bit-write reduction
  in non-volatile memory technologies,'' in \emph{NANOARCH}, 2014.

\bibitem{Farbeh-16-FloatingECC}
H.~Farbeh \emph{et~al.}, ``Floating-ecc: Dynamic repositioning of error
  correcting code bits for extending the lifetime of stt-ram caches,''
  \emph{TonC}, 2016.

\bibitem{Ferreron-15-Concertina}
A.~Ferreron \emph{et~al.}, ``Concertina: Squeezing in cache content to operate
  at near-threshold voltage,'' \emph{Transactions on Computers}, 2015.

\bibitem{Korgaonkar-18-DensityTradeoffs-WCAB-VHC}
K.~Korgaonkar \emph{et~al.}, ``Density tradeoffs of non-volatile memory as a
  replacement for sram based last level cache,'' in \emph{ISCA}, 2018.

\bibitem{Natsui-20-DualPortSOT-MRAM}
M.~Natsui \emph{et~al.}, ``Dual-port sot-mram achieving 90-mhz read and 60-mhz
  write operations under field-assistance-free condition,'' \emph{JSSC}, 2020.

\bibitem{Pekhimenko-12-BaseDeltaImmediate}
G.~Pekhimenko \emph{et~al.}, ``Base-delta-immediate compression: Practical data
  compression for on-chip caches,'' in \emph{PACT}, 2012.

\bibitem{Schechter-10-UseECPnotECC}
S.~Schechter \emph{et~al.}, ``Use ecp, not ecc, for hard failures in resistive
  memories,'' \emph{ACM SIGARCH Computer Architecture News}, 2010.

\bibitem{Wang-13-I2WAP}
J.~Wang \emph{et~al.}, ``i2wap: Improving non-volatile cache lifetime by
  reducing inter-and intra-set write variations,'' in \emph{HPCA}, 2013.

\end{thebibliography}


\end{document}